\documentclass[%
 reprint,%
 amssymb, amsmath,%
 aip,cha,
]{revtex4-1}
\usepackage[columnwise]{lineno}

\usepackage{docs}%
\usepackage{bm}%
\usepackage[colorlinks=true,linkcolor=blue]{hyperref}%
\expandafter\ifx\csname package@font\endcsname\relax\else
 \expandafter\expandafter
 \expandafter\usepackage
 \expandafter\expandafter
 \expandafter{\csname package@font\endcsname}%
\fi
\hypersetup{
	colorlinks=true,
	linkcolor=blue,
	filecolor=magenta,}
\usepackage{graphicx}
\usepackage{lipsum}

\begin{document}

\title{Effect of random anisotropy in stabilization of topological chiral textures} %

\author{Gajanan Pradhan}%
\affiliation{Laboratory for Nanomagnetism and Magnetic Materials (LNMM), School of Physical Sciences, National Institute of Science Education and Research (NISER), HBNI, P.O.- Jatni, 752050, India}%

\author{Subhankar Bedanta}%
\email{sbedanta@niser.ac.in}
\affiliation{Laboratory for Nanomagnetism and Magnetic Materials (LNMM), School of Physical Sciences, National Institute of Science Education and Research (NISER), HBNI, P.O.- Jatni, 752050, India}%
\date{March 2020}%


\begin{abstract}
	Ever increasing demand of skyrmion manipulation in nanodevices has brought up interesting research to understand the stabilization of these topologically protected chiral structures. To understand the actual shape and size of skyrmion observed experimentally, we have performed micromagnetic simulations to investigate skyrmion stabilization in presence of random anisotropy in magnetic thin film system. Previous experimental reports of skyrmion imaging in thin films depicts that the skyrmion shape is not perfectly circular. Here we show via simulations that the shape of a skyrmion can get distorted due to the presence of different local anisotropy energy. The values of uniaxial anisotropy constant ($K_{u}$) and random aniostropy constant ($K_{r}$) are varied to understand the change in shape and size of a skyrmion and an antiskyrmion stabilized in a square magnetic nanoelement. The skyrmion shape gets distorted and the size gets constant for high random anisotropy energy in the system.

\end{abstract}

\maketitle
 Magnetic skyrmions \cite{skyrme1962unified} are non-trivial particle-like configuration of spins having topological protection. These chiral magnetic textures were initially discovered in bulk ferromagnets having non-centrosymmetry such as MnSi and (FeCo)Si\cite{muhlbauer2009skyrmion,yu2010real} and further in ferromagnetic thin films having broken inversion symmetry\cite{fert2013skyrmions,jiang2015blowing,hrabec2017current}. The nucleation and propagation of an individual skyrmion in a nanotrack of magnetic thin films\cite{jiang2015blowing,woo2016observation,hrabec2017current} incorporates the future importance of skyrmions in technological advancements such as high density magnetic storage devices\cite{kiselev2011chiral}. Formation of skyrmions\cite{roessler2006spontaneous, heinze2011spontaneous,schulz2012emergent,seki2012observation,nagao2013direct,nagaosa2013topological,milde2013unwinding} in thin films is prominently due to the presence of Dzyaloshinskii-Moriya interaction (DMI) \cite{dzyaloshinsky1958thermodynamic,moriya1960anisotropic} in the system. Interfacial DMI (iDMI) is an anti-symmetric exchange interaction which occurs at the interface between a heavy metal (e.g. Pt, Ir, Ta) and a ferromagnet (e.g. Co, Fe, CoFeB) due to the presence of large spin orbit coupling and inversion symmetry breaking\cite{fert2013skyrmions,sampaio2013nucleation,iwasaki2013current}. Another special type of skyrmion known as the antiskyrmion was first predicted to exist in systems having bulk DMI\cite{bogdanov1989thermodynamically,bogdanov2002magnetic} and has recently been observed in Heusler compounds with $D_{2d}$ symmetry at room temperature\cite{nayak2017magnetic}. It has also been predicted that magnetic thin films having anisotropic interfacial DMI with $C_{2v}$ symmetry like Au/Co/W films\cite{camosi2017anisotropic} can also host potential antiskyrmions\cite{gungordu2016stability,hoffmann2017antiskyrmions}. The two dimensional spin configuration of a skyrmion can be viewed as a hedgehog model in 3D picture. The skyrmion is defined by a quantized topological number of value $\pm1$ which states that the spins wrap a unit sphere for one time.
 
 For a ferromagnetic system, the total energy ($E_{total}$) constitutes contribution from anisotropy of the system ($E_{anis}$), exchange interaction ($E_{ex}$), demagnetization ($E_{demag}$) and DMI ($E_{DMI}$). It is expressed as
 
 \begin{equation}
 \begin{split}
 E_{total} &= E_{anis}+E_{ex}+E_{demag}+E_{DMI} \\
 &= K_{u}(\hat{\textbf{s}_{i}} . \hat{\textbf{z}})^2 -J_{ij}(\hat{\textbf{s}_{i}} . \hat{\textbf{s}_{j}}) - \frac{1}{2}\mu_{0}\textbf{H}_{demag}.\textbf{M} \\
 &+ \vec{d}_{ij} . (\hat{\textbf{s}_{i}} \times \hat{\textbf{s}_{j}})
 \end{split}
 \end{equation}
 
 where $K_{u}$ is the uniaxial anisotropy constant, $J_{ij}$ and $\vec{d}_{ij}$ are the exchange constant and the DM interaction vector, respectively, between $i^{th}$ and $j^{th}$ spin, $\textbf{H}_{demag}$ is the demagnetization field, $\textbf{M}$ is the magnetization, $\hat{\textbf{s}_{i}}$ represents the atomic moment unit vector for $i^{th}$ site, and $\hat{\textbf{z}}$ is the global easy axis direction\cite{getzlaff2007fundamentals}. The Landau-Lifshifz-Gilbert (LLG) equation given by
 
 \begin{equation}
 \frac{d\textbf{m}}{dt} = -|\gamma|\textbf{m}\times \textbf{H}_{eff} + \alpha \left(\textbf{m}\times \frac{d\textbf{m}}{dt}\right)
 \end{equation}
 
 is solved to minimize the energy of a ferromagnetic system and obtain a groundstate configuration of spins\cite{gilbert2004phenomenological}. Here, $\textbf{H}_{eff} = -1/(\mu_{0}M_{s})(\partial E_{total}/\partial\textbf{m})$, is the effective field generated considering all the interactions, $\textbf{m}$ is the magnetization vector, $\alpha$ is the Gilbert damping constant and $\gamma$ is the gyromagnetic ratio.

\begin{figure}[h!]
	\centering
	\includegraphics[width=0.8\linewidth]{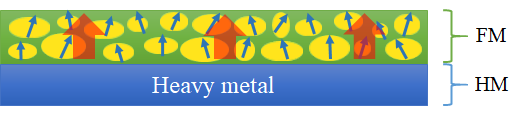}
	\caption{Schematic showing the random anisotropy behaviour in a magnetic thin film system. The large red arrows show the overall uniaxial anisotropy and the small blue arrows show the dispersion in anisotropy giving rise to the  random anisotropy in the thin film.}
	\label{fig:fig-1}
\end{figure}

Skyrmion generation in magnetic thin films is usually observed in systems with perpendicular magnetic anisotropy (PMA). The thickness of the magnetic layer needs to be near the spin reorientation transition (SRT) thickness so that the anisotropy of the system is low. The deposition of a thin film occurs by continuous agglomeration of grains on a substrate. These grains differ in size, texture and morphological characteristics. The magnetic moment of individual grains may point in different direction due to the presence of different local magnetic anisotropies. The spatial variation in local anisotropies can bring sizeable effect in the stablization of magnetic textures\cite{chowdhury2018360,alben1978random,idigoras2011collapse}. Recently we have discussed the importance of random anisotropy in the formation of 360$^{\circ}$ domain walls in a magnetic thin film.\cite{chowdhury2018360} Further we have shown that a comnbination of random anisotropy, dipolar interactions and uniaxial anisotropy it is possible to observe layer-by-layer magnetization reversal in a ferromagnetic/non-magnetic/ferromagnetic system.  \cite{chowdhury2020}

Experimentally, the quantification of the randomness in local anisotropies is very much challenging so far. The schematic of randomness of local anisotropy in magnetic thin film system is shown in  Figure 1. The thin ferromagnetic (FM) layer is made of grains as represented by yellow circles. The individual easy axis orientation of these grains is shown by blue arrows which indicates the dispersion in local anisotropies in the system. The overall uniaxial anisotropy in the system is due to the contribution of all these local alignment of magnetic moments as indicated by the red arrows.

The random anisotropy model was first proposed by Alben et al.\cite{alben1978random} for amorphous ferromagnetic materials. The average anisotropy constant for a thin ferromagnet with grains having randomly oriented easy axis is stated as 

\begin{equation}
\langle K_{1} \rangle = |K_{1}|. \left( \frac{D}{L_{0}} \right)^{6}
\end{equation}

where $K_{1}$ is the local magnetic anisotropy constant, $D$ is the average size of the grains and $L_{0}$ is the ferromagnetic exchange length\cite{alben1978random,herzer1990grain}. The model considers that the  magnitude of local anisotropy for each grain is same for each site. However, the spatial dispersion arises due to difference in direction of local anisotropies. The model also assumes that $L_{0}$ is greater than $D$. The net energy density ($E_{total}$) given in Eq. (1) considering random orientation of local anisotropies, will be modified to 

\begin{equation}
\begin{split}
E_{total} &= E_{anis}+E_{ex}+E_{demag}+E_{DMI} \\
&= K_{u}(\hat{\textbf{s}_{i}} . \hat{\textbf{z}})^2 + K_{r_{i}}(\hat{\textbf{s}_{i}} . \hat{\textbf{r}_{i}})^2 -J_{ij}(\hat{\textbf{s}_{i}} . \hat{\textbf{s}_{j}})\\ 
& -\frac{1}{2}\mu_{0}\textbf{H}_{demag}.\textbf{M} + \vec{d}_{ij} . (\hat{\textbf{s}_{i}} \times \hat{\textbf{s}_{j}})
\end{split}
\end{equation}

where $K_{r_{i}}$ denotes the random anisotropy constant for $i^{th}$ grain and $\hat{\textbf{r}_{i}}$ denotes the local easy axis direction. The LLG equation in our simulation is solved based on the total energy given in Eq. (4). The stability of a chiral spin structure is greatly affected as the system attains the minimum energy state. In this paper, we have tried to understand the effect of random anisotropy for perpendicularly anisotropic samples having interfacial DMI. Micromagnetic simulation provides an unique advantage to quantify the random anisotropy along with the uniaxial anisotropy in a system. The value of uniaxial anisotropy constant ($K_{u}$) and random anisotropy constant ($K_{r}$) are varied and the shape and size of stable spin configurations of skyrmion and antiskyrmion are observed in detail.

\begin{figure}[h!]
	\centering
	\includegraphics[width=1\linewidth]{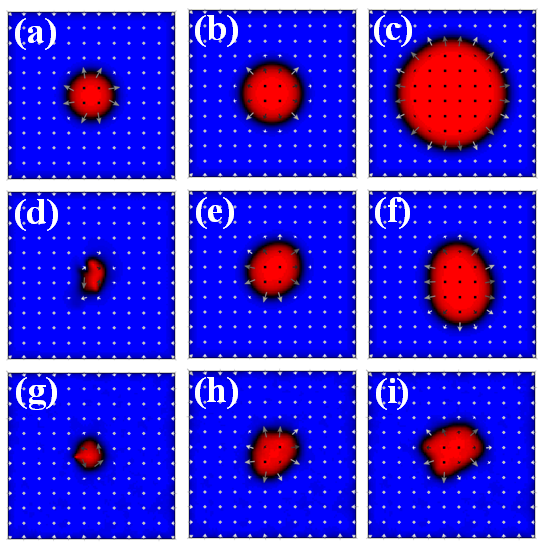}
	\caption{Formation and evolution of a Neel skyrmion from initial ferromagnetics state for three different cases. In case I (a)-(c), $K_{u} = 0.8$ MJm$^{-3}$ and $K_{r} = 0$ MJm$^{-3}$. In case II (d)-(f), $K_{u} = 0.8$ MJm$^{-3}$ and $K_{r} = 0.1$ MJm$^{-3}$ ($K_{r} \ll K_{u}$). In case III (g)-(i), $K_{u} = 0.8$ MJm$^{-3}$ and $K_{r} = 0.8$ MJm$^{-3}$ ($K_{r}=K_{u}$).}
	\label{fig:fig-2}
\end{figure}

The three dimensional micromagnetic simulations were performed using Object Oriented Micromagentic Framework (OOMMF) software developed at National Institute of Standards and Technology (NIST)\cite{donahue1999oommf}. In order to incorporate interfacial Dzyaloshinskii-Moriya interaction (iDMI) in the system, an extension module\cite{rohart2013skyrmion} was used in the simulations. The simulations were performed for a 1 nm thick square nanoelement over an area of 200 nm $\times$ 200 nm. The value of Gilbert damping constant ($\alpha$) is 0.3 and the Gilbert gyromagnetic ratio ($\gamma$) is $-2.211 \times 10^{5}$ m A$^{-1}$ s$^{-1}$. The saturation magnetization ($M_{s}$) is taken to be 580 kA m$^{-1}$, the exchange stiffness ($A$) is 15 pJm$^{-1}$ and the DMI constant ($D$) to be 3.8 mJm$^{-2}$. In order to understand the stabilization of a skyrmion, isotropic DMI has been considered whereas for antiskyrmion, anisotropic DMI has been used. The cell size was kept at 2 nm $\times$ 2 nm $\times$ 1 nm. The thickness of the sample is sufficiently smaller than the exchange length $l_{ex} = \sqrt{\frac{A}{K}} \approx$ 4.3 nm. The random anisotropy is incorporated in the system by using \texttt{Oxs\_RandomVectorField} command which chooses a distinct direction for each local anisotropies. This direction is governed by the value of Randomseed value used in the simulations. We have studied five different set of simulations where we have considered the Randomseed value to be 1-5. The LLG equation for magnetization dynamics is solved by Runge-Kutta method (rkf54). Periodic boundary conditions (PBC) have not been used in the simulations. 

Figure 2 shows the formation and stabilization of a Neel skyrmion in a square nanoelement. $C_{\infty v}$ symmetry in DMI has been considered here where the system is invariant under rotations around the z axis and only two non-zero coefficients remain of the DMI tensor ($D_{12}=-D_{21}=D$). The nucleation process involves the application of current with spin polarization of 0.4 along $-\hat{z}$ direction. The initial state is ferromagnetic and a 100-ps-pulse long current pulse with $J = 8 \times 10^{8}$ A cm$^{-2}$ is applied in a disk of radius $r=17.32$ nm at the center of the nanoelement. The system is further relaxed to reach a stable ground state and the simulation stops when $\frac{d\textbf{m}}{dt} < 0.01$ deg is satisfied for all spatial points of the simulation area. The stable state is reached when the tilt in the magnetization vector per unit second is less than 0.01 deg across all area. Any external magnetic field is not applied during the simulations. Presence of DMI in the system leads to continous rotation of spins from $+\hat{z}$ direction (blue colour region) to $-\hat{z}$ direction (red colour region) which makes it a skyrmionic state and not a bubble domain state. The skyrmion size increases or decreases depending on the energy contribution from iDMI and magnetic anisotropy\cite{rohart2013skyrmion,wang2018theory,behera2018size}.

\begin{figure}[h!]
	\centering
	\includegraphics[width=1\linewidth]{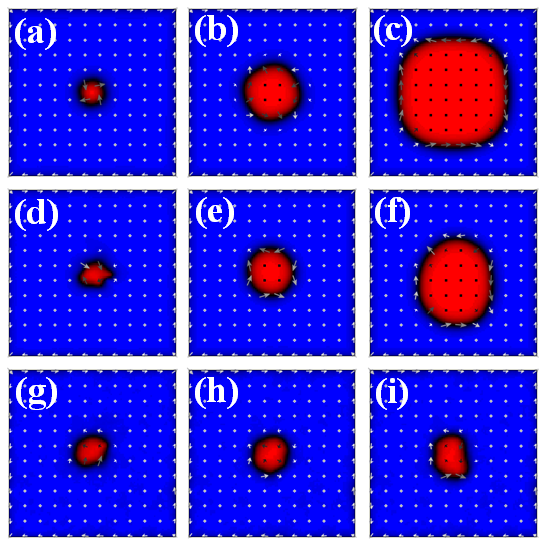}
	\caption{Formation and evolution of an antiskyrmion from initial ferromagnetic state for three different cases. In case I (a)-(c), $K_{u} = 0.8$ MJm$^{-3}$ and $K_{r} = 0$ MJm$^{-3}$. In case II (d)-(f), $K_{u} = 0.8$ MJm$^{-3}$ and $K_{r} = 0.1$ MJm$^{-3}$ ($K_{r} \ll K_{u}$). In case III (g)-(i), $K_{u} = 0.8$ MJm$^{-3}$ and $K_{r} = 0.8$ MJm$^{-3}$ ($K_{r}=K_{u}$).   }
	\label{fig:fig-3}
\end{figure}

The value of random anisotropy constant ($K_{r}$) has been varied keeping the uniaxial anisotropy ($K_{u}$) constant at 0.8 MJm$^{-3}$ to understand the effect of random anisotropy energy on the shape and size of the skyrmion. Randomseed value has been taken to be 5. Three cases have been considered here: Case I: Absence of random anisotropy ($K_{u} = 0.8$ MJm$^{-3}$ and $K_{r} = 0$ MJm$^{-3}$), Case II: Magnitude of random anisotropy is significantly less than uniaxial anisotropy ($K_{u} = 0.8$ MJm$^{-3}$ and $K_{r} = 0.1$ MJm$^{-3}$) and Case III: Magnitude of random anisotropy is same as that of uniaxial anisotropy ($K_{u} = 0.8$ MJm$^{-3}$ and $K_{r} = 0.8$ MJm$^{-3}$). Figure 2(a)-(c) show the evolution into a skyrmionic state from an initial ferromagnetic state for case I. The skyrmion size increases and reaches an ideal skyrmionic state. In case II, shown in figure 2(d)-(f), the skyrmion size also increases. The change in size of the skyrmion in case II is not large as compared to case I since  $K_{r}$ is significantly less than the $K_{u}$. However, the shape of the final stable skyrmionic state is distorted as shown in figure 2(f). This change in shape of the skyrmion is due to the inclusion of local random anisotropy energy in the system. Distortion in shape of a skyrmion has already been reported experimentally\cite{romming2015field,soumyanarayanan2017tunable,romming2013writing,ma2018electric}. This occurs due to inhomogeneity in thin film system like random anisotropy, where local anisotropy directions varies from one site to another. For case III, representated by figure 2(g)-(i), the skyrmion size is remarkably small as compared to other two cases due to high anisotropy.

We have also studied the formation of an antiskyrmion in the same system. However, for antiskyrmion stabilization, $D_{2d}$ symmetry in DMI is considered which is anisotropic and also allows only two non-zero tensor coefficients ($D_{12}=D_{21}=D$). Again three cases have been considered here: Case I: Absence of random anisotropy ($K_{u} = 0.8$ MJm$^{-3}$ and $K_{r} = 0$ MJm$^{-3}$), Case II: Magnitude of random anisotropy is significantly less than uniaxial anisotropy ($K_{u} = 0.8$ MJm$^{-3}$ and $K_{r} = 0.1$ MJm$^{-3}$) and Case III: Magnitude of random anisotropy is same as that of uniaxial anisotropy ($K_{u} = 0.8$ MJm$^{-3}$ and $K_{r} = 0.8$ MJm$^{-3}$). Randomseed value has been taken to be 5. In case I, the size of the antiskyrmion size increases and reaches a final stable state as shown in Figure 3(a)-(c). The antiskyrmion size reduces and the shape gets distorted with slight inclusion of random anisotropy for case II (Figure 3(d)-(f)). The final stable antiskyrmion size for case III is very small compared to other cases due to high anisotropy in the system.

\begin{figure}[h!]
	\centering
	\includegraphics[width=1\linewidth]{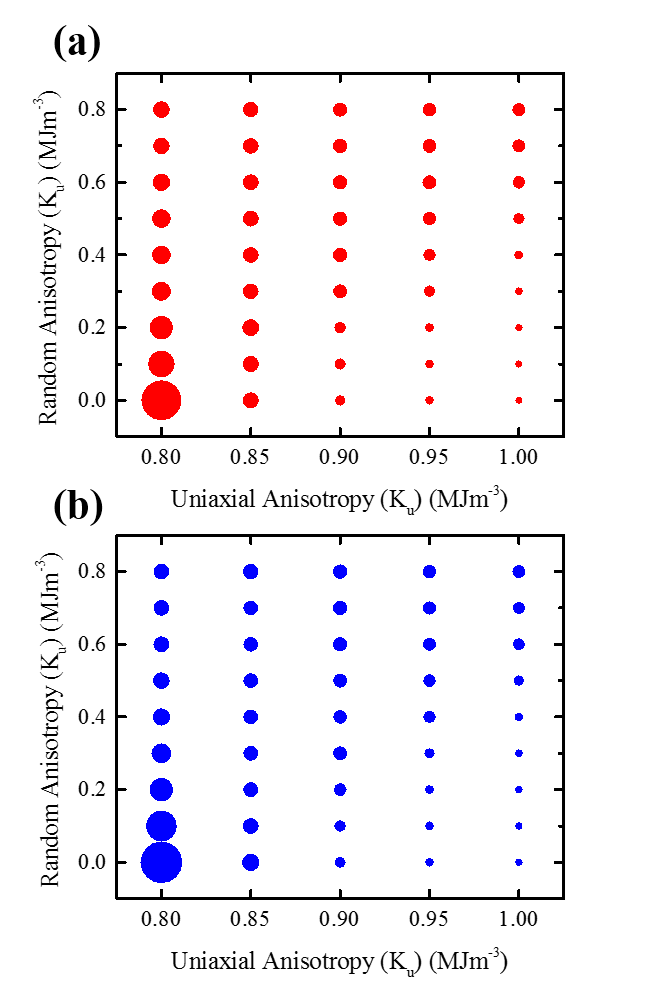}
	\caption{Phase plot showing the effective skyrmion (Plot (a)) and antiskyrmion (Plot (b)) size with variance in Random anisotropy constant (Kr) and Uniaxial anisotropy constant (Ku). The red and blue circular dots represent the relative size of skyrmions and antiskyrmions, respectively, normalized to the maximum size (maximum radius of skyrmion and antiskyrmion are 68.76 nm and 72.13 nm, respectively).}
	\label{fig:fig-4}
\end{figure}

The effective size of the stable ground state of skyrmion (Sk) and antiskyrmion (ASk) was measured for a range of random anisotropy and uniaxial anisotropy energies. The radius of a Sk or an ASk for a particular value of $K_{r}$ and $K_{u}$ is averaged over the values obtained for randomseed = 1,2,3,4 and 5. The value of $K_{r}$ is varied from 0 MJm$^{-3}$ to 0.8 MJm$^{-3}$ and the value of $K_{u}$ is ranged between 0.8 MJm$^{-3}$ and 1.0 MJm$^{-3}$. Figure 4 shows the phase map of skyrmion size (shown in Figure 4(a)) and antiskyrmion size (shown in Figure 4(b)) i.e. the effective radius of the texture at final ground state for a wide range of $K_{r}$ and $K_{u}$ values. The radii of the Sk and ASk for different $K_{r}$ and $K_{u}$ are represented by the red and blue circular dots, respectively, whose size vary relatively with the radius in the plot. For each distorted state, a black and white image was extracted via OOMMF where the spins in $+\hat{z}$ and $-\hat{z}$ direction are represented by white and black pixels respectively. Spins having zero or some $-\hat{z}$ component in their direction are also considered as black pixels and spins having $+\hat{z}$ component in their direction are represented by pixels with intermediate colours. Using color threshold method in imageJ software, the intermediate colors between black and white are thresholded to white pixels. The number of black pixels are then counted using histogram. The area of black pixels is then calculated in $nm^2$ and the radius of the chiral texture is formulated.

It is observed that the size of the Sk and ASk decreases with increase in magnitude of the uniaxial anisotropy of the system. The randomness in anisotropy which is incorporated using an extra random anisotropy constant ($K_{r}$), is fixed along a particular direction for each local sites. This particular unit vector representing the direction of random anisotropy can have arbitrary component of $x$, $y$ and $z$ directions. For an ideal case, considering only uniaxial anisotropy, the Sk and ASk are starting to evolve from initial ferromagnetic state representing spins along $+\hat{z}$(blue colour) and $-\hat{z}$(red colour) which are the easy axes directions. In presence of an extra anisotropy i.e. random anisotropy having comparable strength ($0.6 - 0.8 MJm^{-3}$) with the uniaxial anisotropy along some other direction, the Sk and ASK sizes do not vary much with increase in uniaxial anisotropy as shown in the plot. 
  
The stability of a skyrmion from an initial ferromagnetic state in the presence of random anisotropy has been studied here via micromagnetic simulations. The size and shape of a final stable skyrmionic state is largely affected by the inclusion of localized random anisotropy energy in the system. In a real thin film system with PMA, the grains have slightly different local anisotropic directions. However, it is very challenging to quantify the value of anisotropy value for each local sites. Micromagnetic simulations provides us a great advantage to understand the effect of random anisotropy on skyrmion statics. In previous experimental observations of a skyrmion, it has to be noted that the skyrmion shape is not perfectly symmetric. In our simulations, we have elucidated that considering random anisotropy in a system the shape of a skyrmion is getting distorted. The values of $K_{r}$ and $K_{u}$ have been varied within experimental feasible range and the stabilization of Sk and ASk is studied. The size of the Sk and ASk reduces with increase in $K_{r}$ for low uniaxial anisotropic samples ($K_{u} < 0.85$ MJm$^{-3}$). For high anisotropic samples, the Sk and ASk size  increases when random anisotropy is added in the system. From the phase study of Sk and ASk size, it is seen that when the random anisotropy is very high in system, the change is skyrmion size is not noticeable. Further, the size and shape distortion of a skyrmion may depend on the percentage of randomness in anisotropy present in a system. Therefore it is important to look at this aspect from experimental point of view. Attention must be put to prepare thin films with less roughness and homogeneous growth so that the random anisotropy will be minimal. By minimizing the random anisotropy we can get the skyrmion shape to be circular to avoid any spurious effects which hinders technological applications.

\bibliography{references}
\end{document}